# Video Analysis and Modeling Performance Task to promote becoming like scientists in classrooms

**Loo Kang Wee[1], <u>Tze Kwang Leong</u>[2]**

lawrence_WEE@moe.gov.sg, tzekwang.LEONG@rgs.edu.sg

[1]*Ministry of Education, Educational Technology Division, Singapore.*
[2]*Ministry of Education, Raffles Girl's School, Singapore*.

## Abstract

This paper aims to share the use of Tracker—a free open source video analysis and modeling tool—that is increasingly used as a pedagogical tool for the effective learning and teaching of Physics for Grade 9 (Secondary 3) students in Singapore schools to make physics relevant to the real world. We discuss the pedagogical use of Tracker, guided by the Framework for K-12 Science Education by National Research Council, USA to help students to be more like scientists.  For a period of 6 to 10 weeks, students use a video analysis coupled with the 8 practices of sciences such as 1. Ask question, 2. Use models, 3. Plan and carry out investigation, 4. Analyse and interpret data, 5. Use mathematical and computational thinking, 6. Construct explanations, 7. Argue from evidence and 8. Communicate information.

This paper's  focus in on  discussing some of the performance task design ideas such as 3.1 flip video, 3.2 starting with simple classroom activities, 3.3 primer science activity, 3.4 integrative dynamics and kinematics lesson flow using Tracker progressing from video analysis to video modeling, 3.5 motivating performance task, 3.6 assessment rubrics and lastly 3.7 close mentorship.

Initial research findings using pre- and post- perception survey, triangulated with student interviews suggest an increased level of students' enjoyment such as "I look forward to physics lessons", "I really enjoy physics lessons" and "Physics is one of the most interesting school subjects" etc for the more mathematically inclined  students. Most importantly, the artefacts of the students' performance task in terms of the research report and Tracker *.TRZ files, further suggest that the use of the Tracker for performance tasks, guided by the Framework for K-12 Science Education by National Research Council, USA, can be an innovative way to mentor authentic and meaningful learning that empowers students to be more like scientists.

# 1    Background

We make reference to the 2012 Framework for K-12 Science Education by National Research Council, USA[1] for the approach in video analysis[2, 3] and modeling[4, 5] to allow students to be like scientists. This conceptual framework aims to capture students' interest and provide them with the necessary foundational knowledge in the field of science and engineering. In short, we empower students to become like scientists in classrooms by mentoring them to 1. Ask question, 2. Use models, 3. Plan and carry out investigation, 4. Analyse and interpret data, 5. Use mathematical and computational thinking, 6. Construct explanations, 7. Argue from evidence and 8. Communicate information.

# 2    Problem

| *Traditional Lessons* | *Scientist Research* |
|---|---|
| Topics taught in isolation | Knowledge from various topics are required |
| Simplified theoretical scenario with many assumptions | Authentic collected data which often include anomaly |
| Knowledge apply to assessment questions | Knowledge apply to real world situations |
| Teacher directed; | Student directed; |
| Teacher decide on question | Ownership of research question |
| No differentiation; One size fits all | Research is individualized |

**Table 1**: Comparison of traditional lessons versus the scientist research

Table 1 summarises the general weaknesses in traditional lessons and the strengths of the scientist research approach in allowing students to experience[6] physics phenomena of the student's choice. Without going into the details of table 1, we aimed to get our students use knowledge from various topics in Physics, experience authentic data collection, apply knowledge to real world situations, students self-direct their learning thus having greater ownership of their own research question and experience individualised research.

# 3    Approach

## 3.1    Flip Video

A series of YouTube instructional videos were created by the authors to demonstrate simpler tasks such as a) Installing Tracker [2, 7-12]  b) Generate displacement $x$ or $y$ direction vs time $t$, velocity $v$ vs time $t$ and acceleration $a$ vs time $t$ graphs and c) Analyse graphs by finding gradient[1] and area[2]. In addition, the flip video also introduced Tracker's interface to students at their leisure time at home, to promote richer discussions of the Physics in the classroom later.

## 3.2   Starting with Simple Classroom Activities

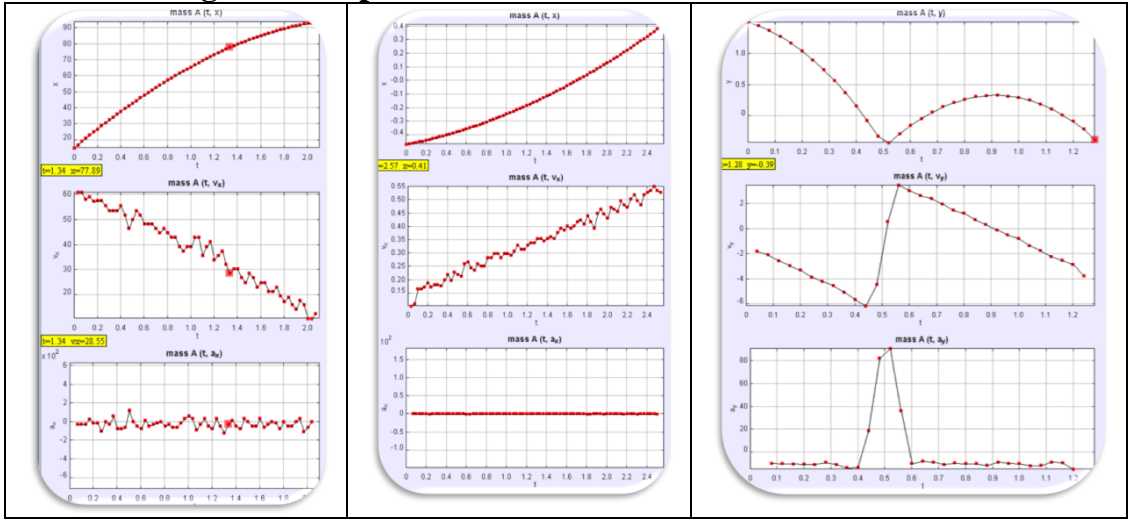

Figure 1.   **LEFT to RIGHT** a) Cart push on level slope slowing down at constant negative acceleration) b) Cart on slope speeding up at constant acceleration and c) Bouncing ball, complex accelerating motion divided into phases.

Three simple video analyses (Figure 1) were provided to the groups of 3 to 4 students, where they discussed, presented and critiqued the other groups' presentations. The selected simple videos are a) cart push on level slope slowing down at constant negative acceleration) b) cart on slope speeding up at constant acceleration and c) bouncing ball [3], complex accelerating motion divided into phases. We found the introduction of simple video to be effective in getting students to relate Physics in the Tracker software as oppose to more complex motion to reduce the cognitive loading.

## 3.3   Primer Scientist Activity

We also conducted a beginner-primer hands-on activity to prepare students for the performance task. The suggested set of equipments are 2 marbles, a ramp,

---

[1] https://www.youtube.com/watch?v=H_zrkl16BNs
[2] https://www.youtube.com/watch?v=7_TgOSMqRQs

and a metre rule where students perform practices of scientists scaffolded by a smaller set of K-12 Science Education (Figure 2) such as

a. Practice 1: Define a problem or ask a question

b. Practice 3: Plan out an investigation

c. Practice 4: Analyse and interpret the data

d. Practice 6: Construct explanations

As an indication of the success of this primer scientist activity, we share some of the more interesting activity-questions asked by the students include experiments such as 1) A marble rolling up and down a ramp, 2) A bouncing marble down stairs or slope, 3) A projectile marble colliding with wall and 4) two colliding marbles[3].

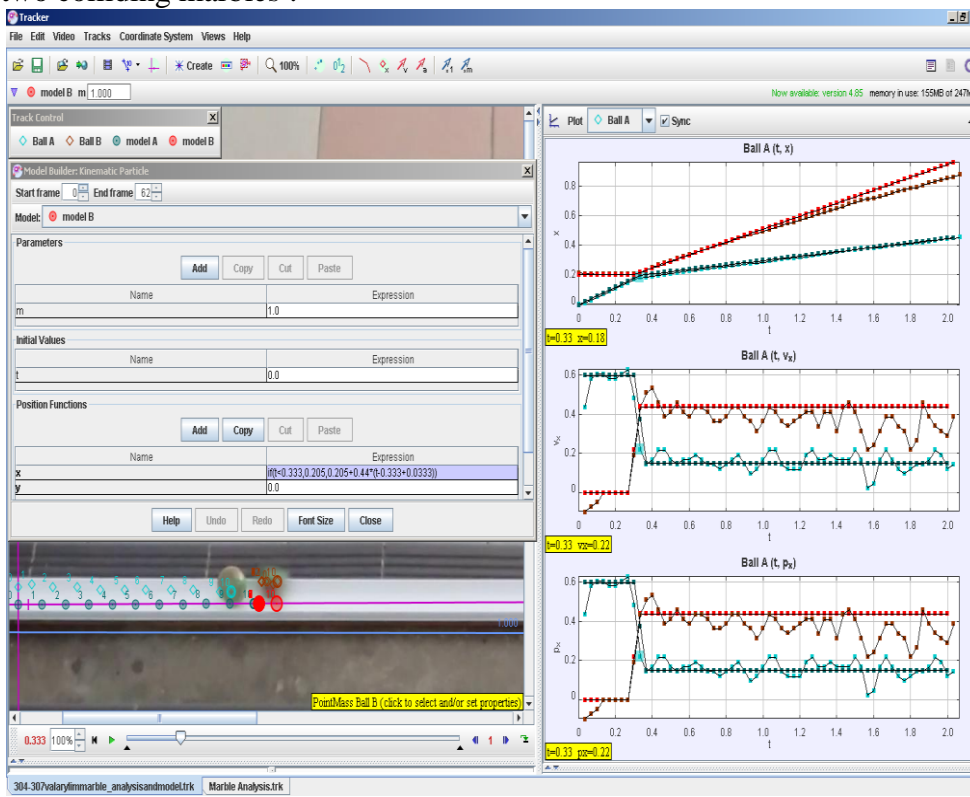

Figure 2.   Students directed video analysis inquiring on the inelastic collision of 2 marbles on a track. Teal trail is the right moving marble and the Red trail is the right moving marble. Model A and B were added on later by the authors for teacher professional network learning purposes.

## 3.4 Integrative Dynamics Lessons from Kinematics

## 3.4.1 Revisit Gentle Push on Horizontal Slope

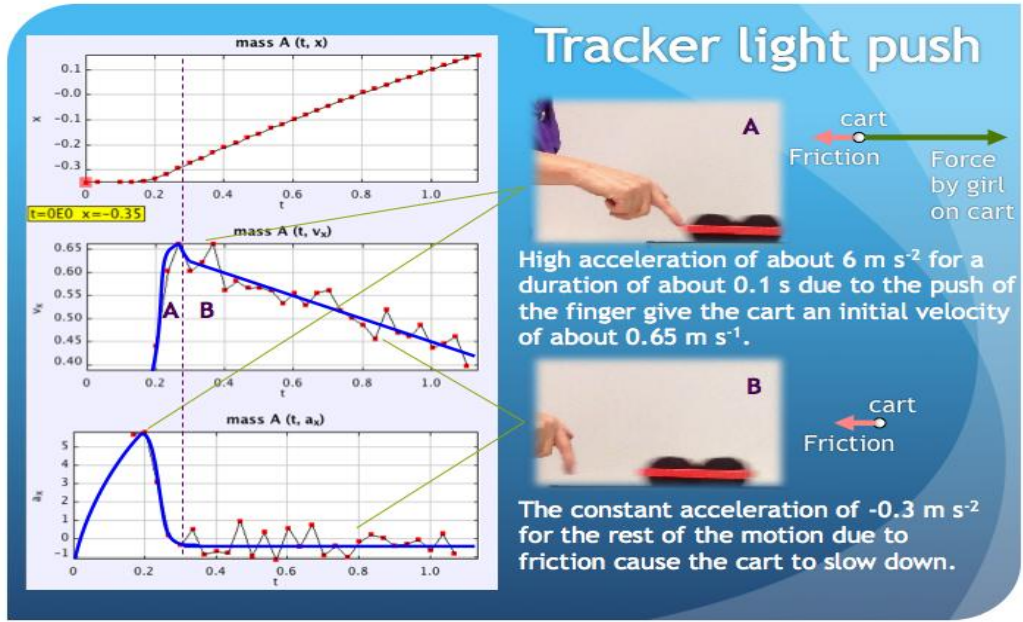

Figure 3. A dynamics lesson using Tracker again to illustrate frictional forces bridging from kinematics topics allows students to build understanding based on what they have already experience themselves in kinematics topics with the now more familiar Tracker horizontal displacement $x$ vs time $t$, horizontal velocity $vx$ vs $t$ and horizontal acceleration $ax$ vs $t$ graphs.

The previous kinematics videos were revisited to discuss the dynamics (Figure 3) where students were asked to suggest possible forces causing the motion. At this stage of the lesson design ideas, we recommend using the Tracker's Dynamics [7-9] model builder to incrementally suggest better models that represent the motion under investigations that has practice 5: mathematical and computational [7] thinking.

### 3.4.1.1 Practice 5: Mathematical Thinking

To model frictional force motion (Figure 4) of the cart moving horizontally, students can mathematically determine the gradient of the horizontal velocity $vx$ vs time $t$ graph to determine acceleration in the $x$ direction, $ax$.

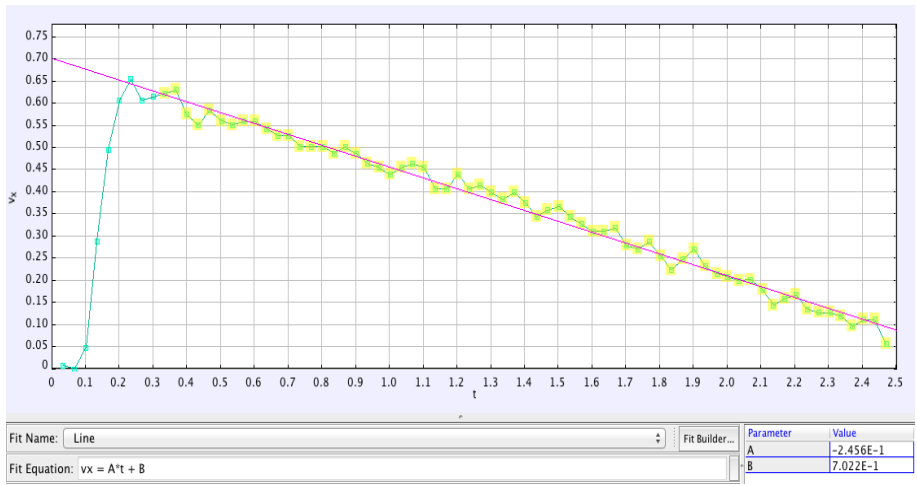

Figure 4.   Tracker DataTool showing a Fit Line of *vx* = -0.2456*t*+0.7022.

### *3.4.1.2   Practice 5: Computational Thinking[13]*

Having determined *ax* and knowing mass of cart to be *m* = 0.2 *kg*, the computational line (Figure 5) —with the "*if*" statement —allows the push force *fx* = 3.2**m* from *t* =0 to 0.2 *s*, and frictional force *fx* = -0.246**m* to be present after time *t* is greater than 0.2 *s*.

$$fx = if\ (t<0.2\ ,\ 3.2*m\ ,\ -0.246*m)$$

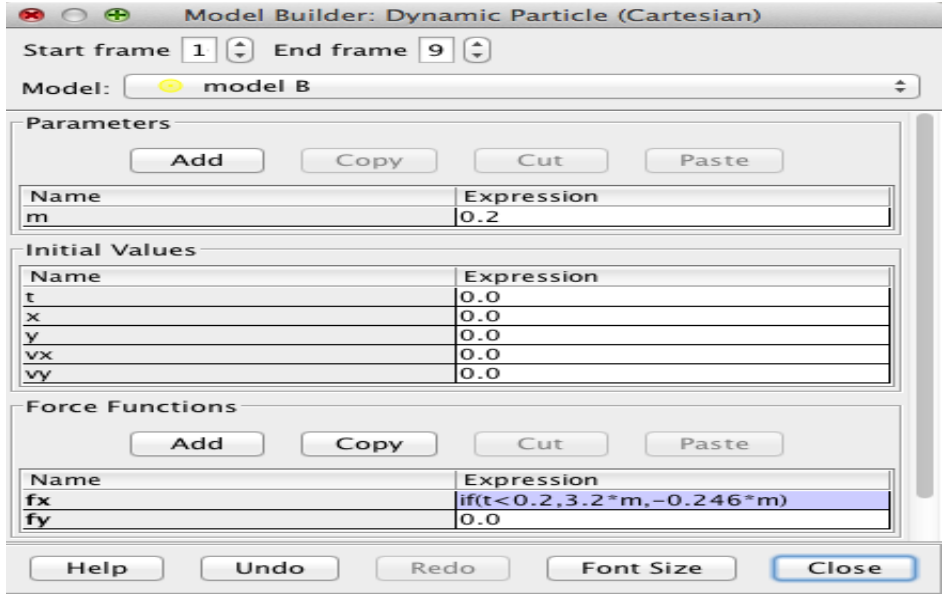

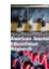

Figure 5. Tracker Dynamics Model Builder interface showing mass $m = 0.2$ $kg$ and force model $fx = if$ (t<0.2,3.2*m,-0.246*m).

## 3.4.2 Atwood Machine

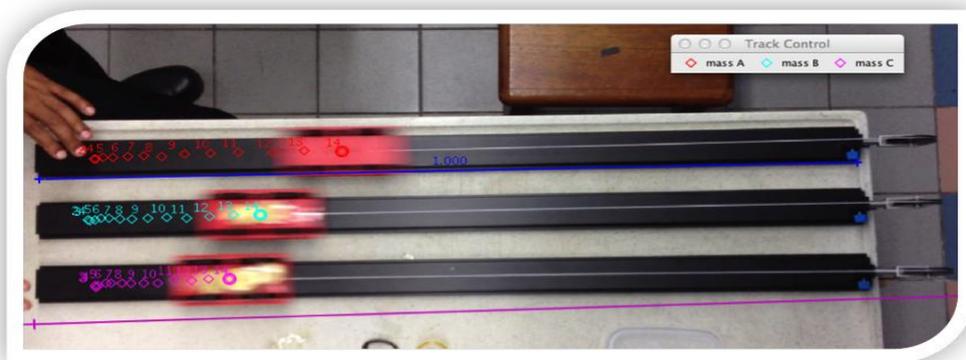

Figure 6. Atwood Machine video showing 3 different mass, $m$ carts, pulled by the same weight $F$ (RIGHT) and their respectively accelerations $ax$.

Another example we tried was the Atwood machine (Figure 6) where 3 carts of different mass were pulled by identical weights. This activity aimed to generate discussions on the relationship between the resultant force $F$ pulling the carts, the carts' acceleration $a$ and the mass of the carts as in Newton's second law.

$$F = ma$$

Again, we recommend linking this video's kinematics motion to the dynamics particle model to provide richer mathematical and computational thinking similar to 3.4.1.1 and 3.4.1.2.

## 3.5 Meaningful Scenario

We gave the students a scenario that simulates the work of real scientists. For example, the following task was assigned:

"You are a scientist who is tasked by A*Star (A local Research initiative) to explain a complex motion and the cause of the motion. You are to record a video of a moving object and to analyze the kinematics and dynamics involved in the motion with the aid of Tracker software. Your report will help the scientific community better understand the complex motion."

## 3.6  Assessment Rubrics (20% of examination score)

| ASSESSMENT RUBRICS ON PERFORMANCE TASK ON ANALYSIS OF MOTION USING VIDEO TRACKER YEAR THREE (RP PHYSICS) | | | | | |
|---|---|---|---|---|---|
| **Level**<br><br>**Criteria** | **Excellent**<br>**(4m)** | **Proficient**<br>**(3m)** | **Adequate**<br>**(2m)** | **Limited**<br>**(1m)** | **Insufficient / No evidence (0m)** |
| **Identify motion to be investigated (C1)** | Identify a motion that is **complex** and **well defined** and involves non-rigid body, multiple objects or multiple phases | Identify a **complex** motion that is **well defined**. | Identify a motion that is **well defined**. | Identify a motion that is **ill-defined**. | |
| **Plan the procedure and filming (C2)** | Use a **comprehensive** and **detailed** procedure to film object in order to ensure **precision** and **accuracy** of measurement. | Use a **clear** and **workable** procedure to film object in order to ensure **precision** and **accuracy** of measurement. | Use a **simplistic** procedure to film object with **some** consideration of **accuracy** of measurement. | Use an **ambiguous** procedure to film object with **little** consideration of **accuracy** of measurement. | |
| **Present graphs with annotation (C3)** | Present graphs **logically** and **clearly** in an appropriate form with relevant **annotation**. | Present graphs **reasonably well** in an appropriate form with relevant **annotation**. | Present relevant but **incomplete** set of graphs | Present graphs **with severe conceptual error**. | |
| **Provide a discussion of the motion (C4)** | Provide a **detailed** and **comprehensive discussion** of the motion. | Provide a **relevant discussion** of the motion with **no errors**. | Provide a discussion of motion with **some minor errors**. | Provide a discussion of motion with **severe conceptual errors**. | |
| **Explain the forces in relation to the motion (C5)** | Explain the forces in relation to the motion by **consistently** making **accurate** inferences from graphs and video. | Explain the forces in relation to the motion by making **some accurate** inferences from graphs | Explain the forces in relation to the motion by making inferences from graphs with **some errors**. | Explain the forces in relation to the motion by with **little attempt** to make inferences from graphs. | |

Table 2: Assessment Rubrics communicating clearly the expected performance indicators of excellent task

We found that the assessment rubrics in Table 2 coupled with the 20% of the examination scores to this performance task, provided students with the extrinsic motivation to self direct their learning [14].

## 3.7   Close Teacher Mentorship

For the learning to go well, students are given opportunities to discuss their analyses with the teachers both in class and outside classroom via weekly consultations. The teachers suggested refinement to their videos which directed them to further readings as well as guided deeper analyses.

# 4   Findings
## 4.1   Students' Pre- and Post- Perception Survey

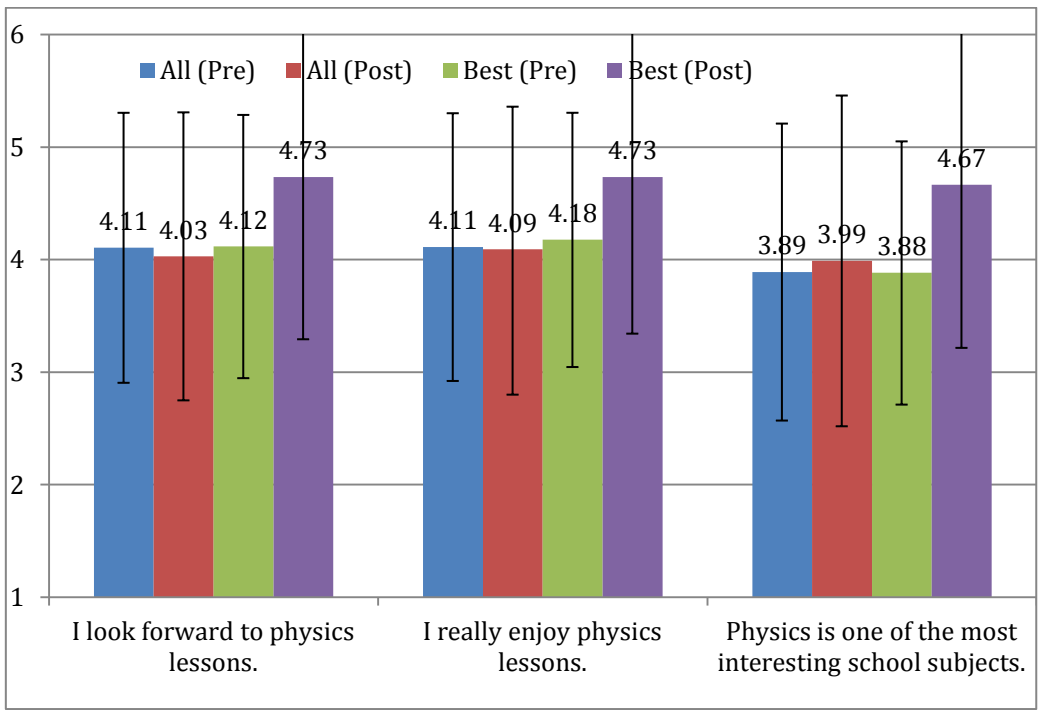

Figure 7.   **N=273** whole cohort of students and **N =30** for high-performing and mathematically-inclined class' **pre- and post-** perception survey on a Likert scale from 1 (strongly disagree) to 6 (strongly agree), middle is 3.5 point. Note the mean and standard deviation are added for ease of interpreting the self reporting perception survey

Our initial analysis of N=273 students' pre- and post- perception survey on the experience of the performance task suggested that only the high-performing and mathematically-inclined class registered a strong positive change (Figure 7) in the affective domains like "I look forward to physics lessons" pre=4.12, post=4.73, "I really enjoy physics lessons" pre =4.18, post = 4.73 and "Physics is one of the most interesting school subjects" pre=3.88, post =4.67 while as a whole cohort of 273 students, there was practically no change in the self-reporting perception survey. We speculate that the no change in pre- and post-perception for the remaining cohort was due two main factors such as 1) student's high perception of themselves like scientists during pre-survey, 2) students were "pushed" too hard to be like scientists in the 6 to 10 weeks, resulting in deliberate low post-survey scores.

## 4.2 Students Actual Artifacts of Performance

We argue that when we evaluate actual student's research report with the Tracker video analysis *.TRZ files, we were able to judge the students' propensity to think like scientists. As every student's experience with the performance task is different since this is largely dependent on the student's individual research question for video analysis, we highlight three acceptably complex performance tasks by three students as a proxy to their becoming scientists' goal in classroom setting.

## 4.2.1 Example: Roller-blading Down an Inclined Slope

From the depth of the analysis in the student's report (Figure 8) coupled with the video analysis, we were pleasantly surprised by the depth of the 4 phases analysis by the student to break a complex motion into simpler parts.

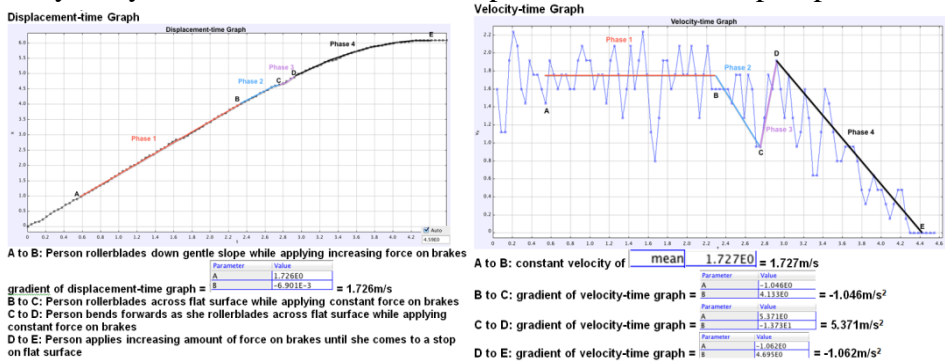

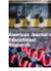

**Figure 8.** Sample screen shots of the students report on the complex roller blading motion broken into simpler 4 phases of motion as determined by the student, mentored by the teachers.

An area of improvement for our next implementation is the stronger emphasis on modeling [8, 13] in the performance task especially if the motion is suitable to be dynamically modeled as in Figure 9.

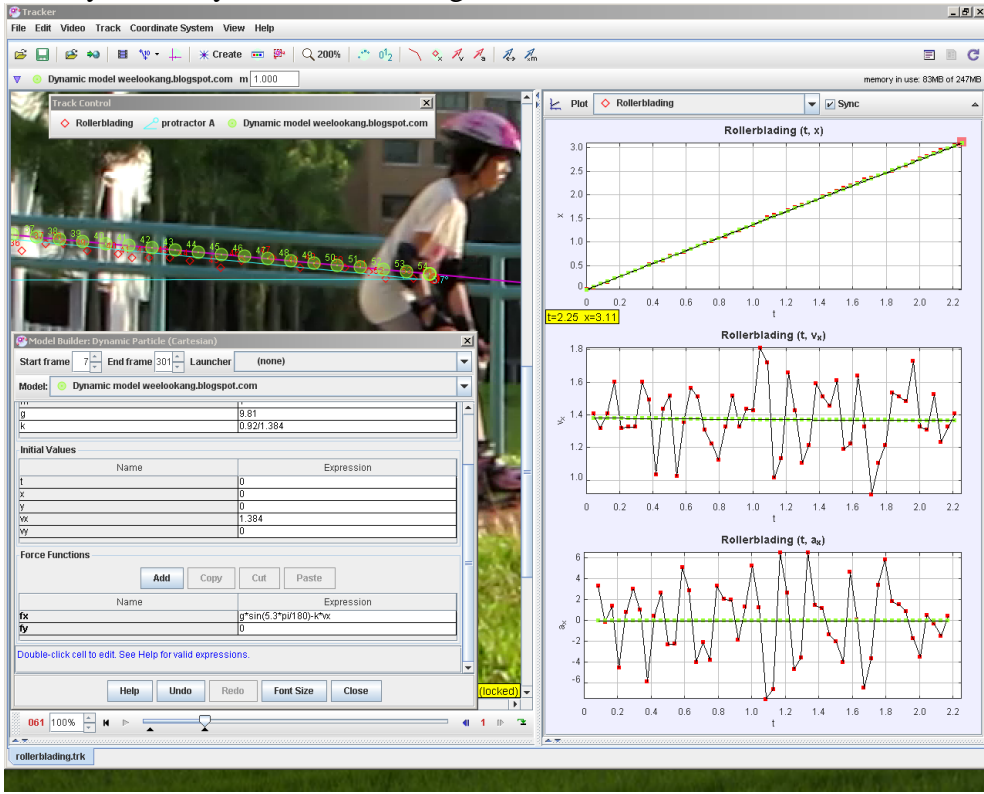

**Figure 9.** roller-blading[4] down an inclined slope video analysis by student with a model added by the teachers to guide-mentor the students where initial velocity, vx' = 1.384, forces in x' and y' parallel and perpendicular to the slope directions are defined as Fx' = g*sin(5.7*pi/180)-k*vx where pi =π = 3.14159, gravitational acceleration, g = 9.81, air drag coefficient k= 0.708 and lastly Fy' = 0

## 4.2.2 Example: Carom Motion with Collision against Wall and Drag Force

In this particular performance task, the student attempted to break the motion of the carom striker into simpler parts and the research report analysis was impressive for a Secondary three student (Figure 10).

---

[4] http://weelookang.blogspot.sg/2014/05/tracker-koaytzemin-student-video-roller.html

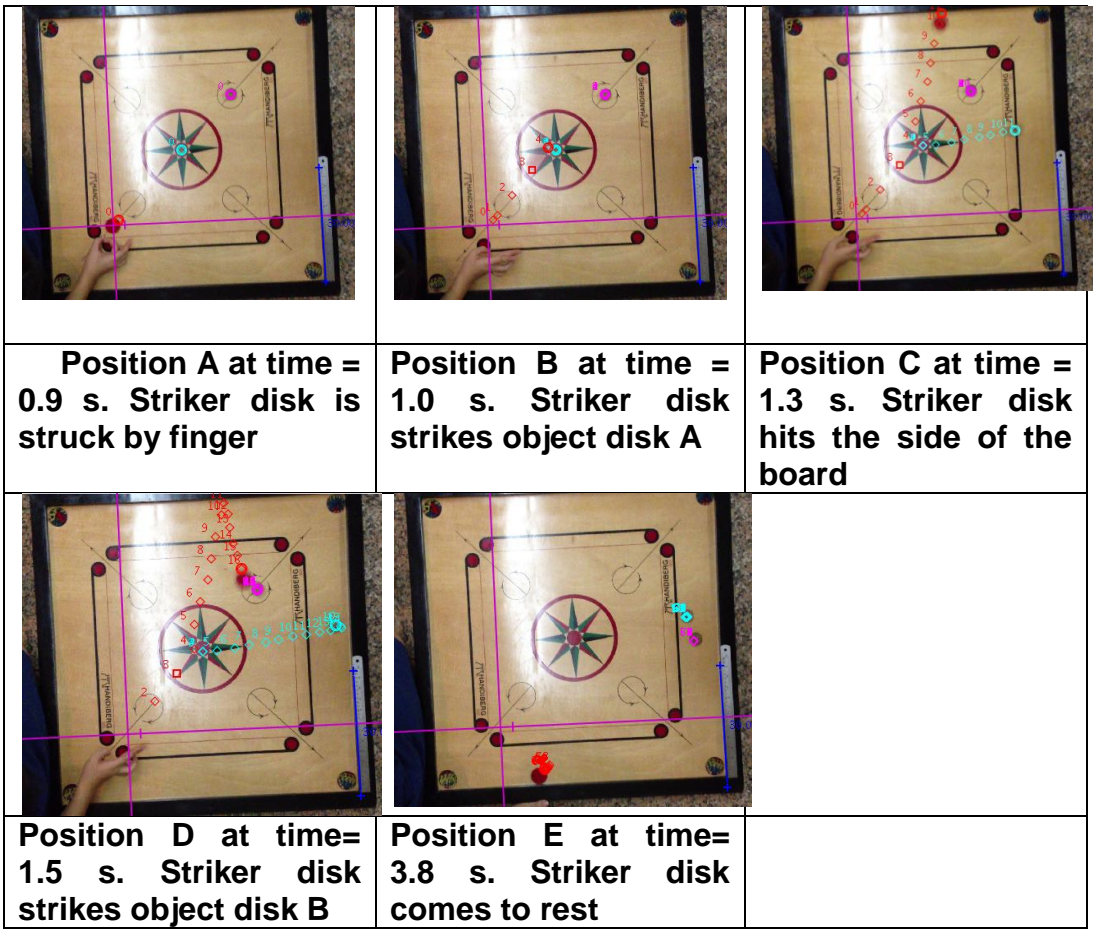

| Position A at time = 0.9 s. Striker disk is struck by finger | Position B at time = 1.0 s. Striker disk strikes object disk A | Position C at time = 1.3 s. Striker disk hits the side of the board |
| --- | --- | --- |
| Position D at time= 1.5 s. Striker disk strikes object disk B | Position E at time= 3.8 s. Striker disk comes to rest | |

Figure 10. Carom stiker motion includes position A at t=0.9 s, striker disk (red) struck by finger, position B at t=1.0 s, striker disk strikes object disk A, position C at time = 1.3 s, striker disk hits the side of the board, position D at time= 1.5 s, striker disk strikes object disk B (magenta) and Position E at time= 3.8 s. Striker disk comes to rest.

We also found the dynamic modeling approach[5] (Figure 11) to be suitable to mentor the student to develop a deeper appreciation of the Physics involved. For example, assuming mass m = 1kg for simplicity, the contact force could be estimated by assuming contact time from t = 0.167 to 0.171 s, of δt = 0.05 s to be contact force on striker disk to be Fx1 = -200 N, Fy1 = -20 N in the Cartesian coordinate system (Figure 11 magenta axes) defined in Tracker. Similarly, contact force with upper wall of carom board can again be estimated by modeling the subsequent motion of the striker disk using contact force with upper wall of carom board to be Fx2 = -40 N, Fy2 = -520 N, assuming time of collision is t =0.458 s to 0.462 s of δt = 0.05 s. Notice the model no longer is able to represent the motion of the striker disk after t > 0.462 s as the model is uniform motion while the real striker disk motion has a weird curve trajectory probably due to the spinning motion.

---

[5] http://weelookang.blogspot.sg/2014/08/carom-collision-force-model.html

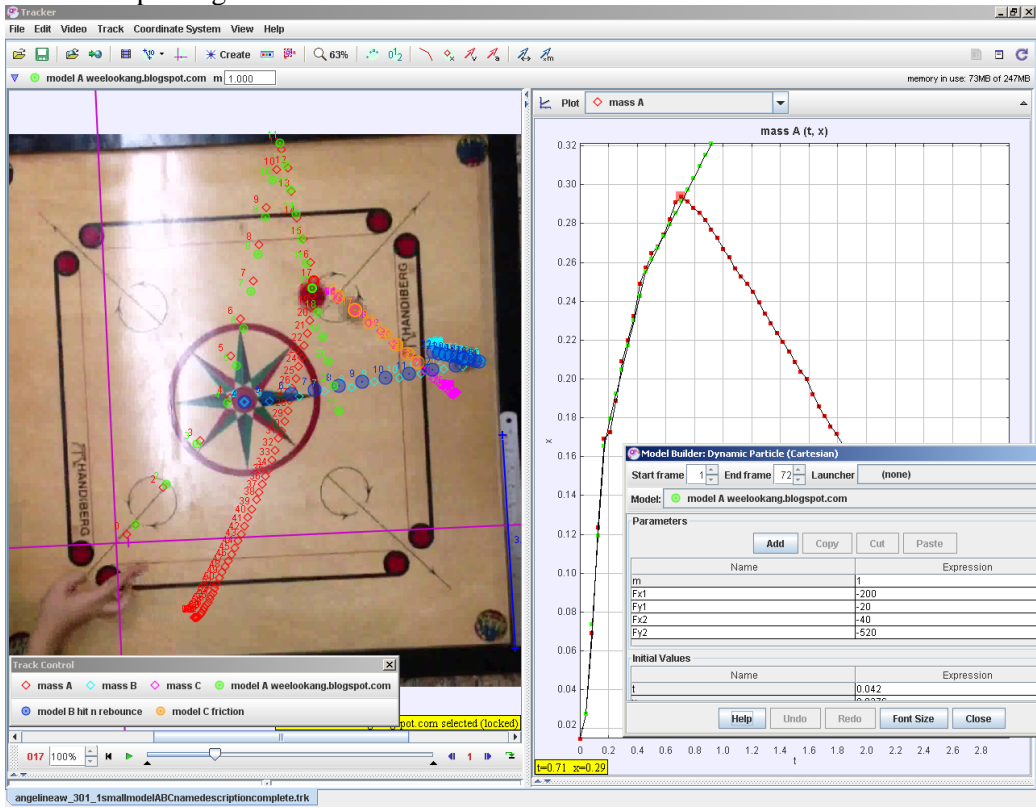

Figure 11. The carom stiker disc motion (red) of motion A-B-C-D is modeled in Tracker model A (green) allowed for deeper mathematical and computational thinking. model A included position A-B-C-D modeled with the following variables contact force with (blue) disk at centre of carom board, Fx1 = -200, Fy1 = -20, contact force with upper wall of carom board Fx2 = -40, Fy2 = -520. Initial values are, time of model t = 0.042, horizontal x position, x = 0.0276, vertical y position y = 0.0252, horizontal velocity vx = 1.103, vertical velocity vy =1.3218. The dynamics model is
fx = if(t<0.167,0,if(t<0.171,Fx1,if(t<0.458,0,if(t<0.462,Fx2,0)))) and
fy = if(t<0.167,0,if(t<0.171,Fy1,if(t<0.458,0,if(t<0.462,Fy2,0))))

## 4.2.3 Example: Balloon Propeled Collision Cars

The final example of students' actual performance task involved the collision of balloon propelled cars. This student's research report includes details of her experimental setup (Figure 12) and we liked the inclusion of the experimental setup investigation and the close-up of the car A and B and the data analysis of velocity in x direction *vx* versus time *t* of the motion of car A moving from left to right.

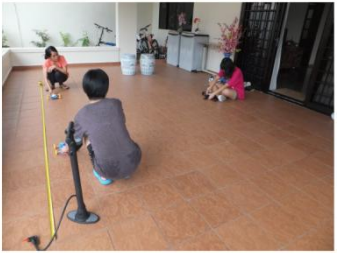

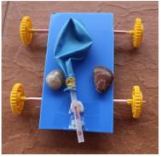

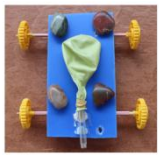

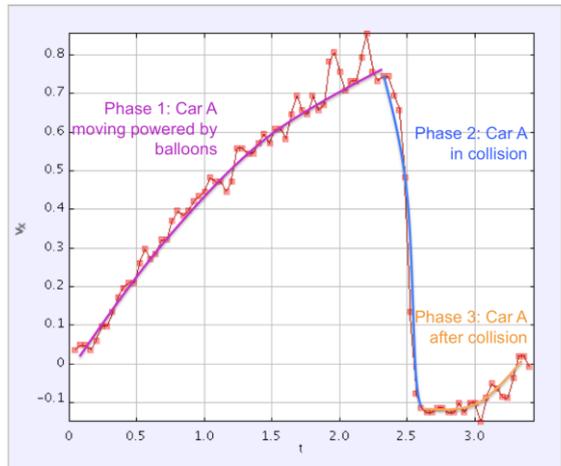

Figure 12. On the left is the experimental setup and the close-up of the Car A and Car B with the balloon deflated. On the right is the data analysis of the motion of Car A moving to the right as it is propeled by the releasing of air from the inflated balloon.

Building on what the student's already analysed, again we highlight the modeling approach[6] to deepen the description of the motion of car A by taking the coefficients of the parabolic curve fit to construct an accelerating model of Car A (Figure 13). We recommend constructing a uniform acceleration motion model A (blue) where the real motion of Car A is not well represented by before suggesting a evidence based model B of a non-uniform acceleration as the balloon deflates non-uniformly, giving the car A a dynamic force in the *x* direction *fx = (-1.853E-1\*t^2+2.551E-1\*t+3.494E-1)\*m* where mass of Car A is found by the students to be *m =0.125 kg*.

---

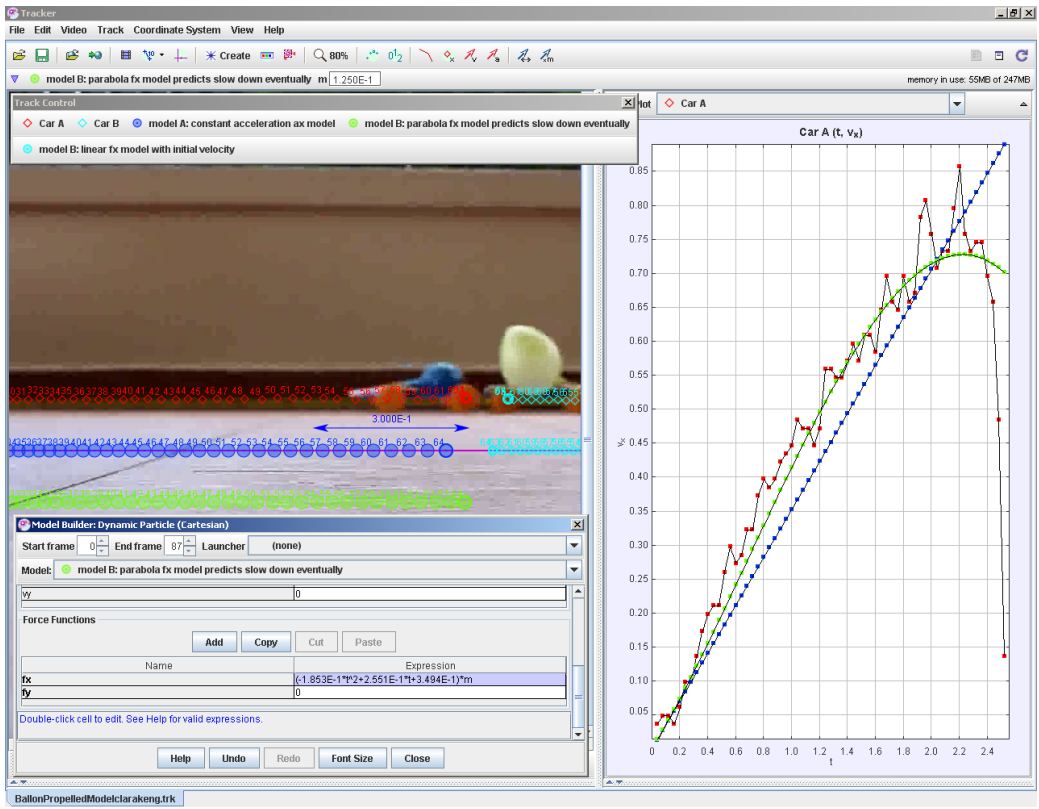

Figure 13. Model A is a constant accelerating model (blue) that does not match the real motion of the balloon propeled car A. Model B is a data analysis evidence based model of dynamic force in the x direction *fx = (-1.853E-1\*t^2+2.551E-1\*t+3.494E-1)\*m* where mass of Car A is found by the students to be *m = 0.125 kg*. Notice how Model A does not fit the real motion whereas the Model B is clearly a better model to describe the real motion of car A.

To summarize, while the pre- and post- perception survey only registered a noticeable increase for the high-performing and mathematically-inclined class (Figure 7), we argue that the examples of the variety and choice of performance tasks, suggests strong indicators of behaving like scientists in classroom context.

We also shared some ideas how to use the dynamic particle model building process in Tracker to further strengthen 2. Use models and 5. Use mathematical and computational thinking.

# 5    Conclusion

This paper describes the use of Tracker—a free open source video analysis and modeling tool—when guided by the Framework for K-12 Science Education by National Research Council, USA can be a powerful approach to make Physics education relevant to the real world and help students experience the 8 key practices of scientists, through a 6 to 10 week performance task.

We discussed some of the performance task design ideas as we believe it will help readers implement their own lessons and they are 3.1 flip video, 3.2 starting with simple classroom activities, 3.3 primer science activity, 3.4 integrative dynamics and kinematics lesson flow using Tracker progressing from video analysis to video modeling, 3.5 motivating performance task, 3.6 assessment rubrics and lastly 3.7 close mentorship.

Initial research findings using pre- and post- perception survey, triangulated with student interviews suggest an increased enjoyment of learning such as I look forward to physics lessons", "I really enjoy physics lessons" and "Physics is one of the most interesting school subjects", for the more mathematically inclined students.

Most importantly, the artefacts (see 4.2.1) of the student's performance task in terms of the research report and Tracker *.TRZ files further suggests majority of the students did find the performance task to be an effective way to mentor authentic and meaningful learning [15] and promoting students to be more like scientists in classrooms.

More students' video analysis and teacher's mentoring models are open educational resources[16], accessible and adaptable through Tracker as a Shared Library [13] (Figure 14) as well as the following URL http://iwant2study.org/lookangejss/ for the benefit of all.

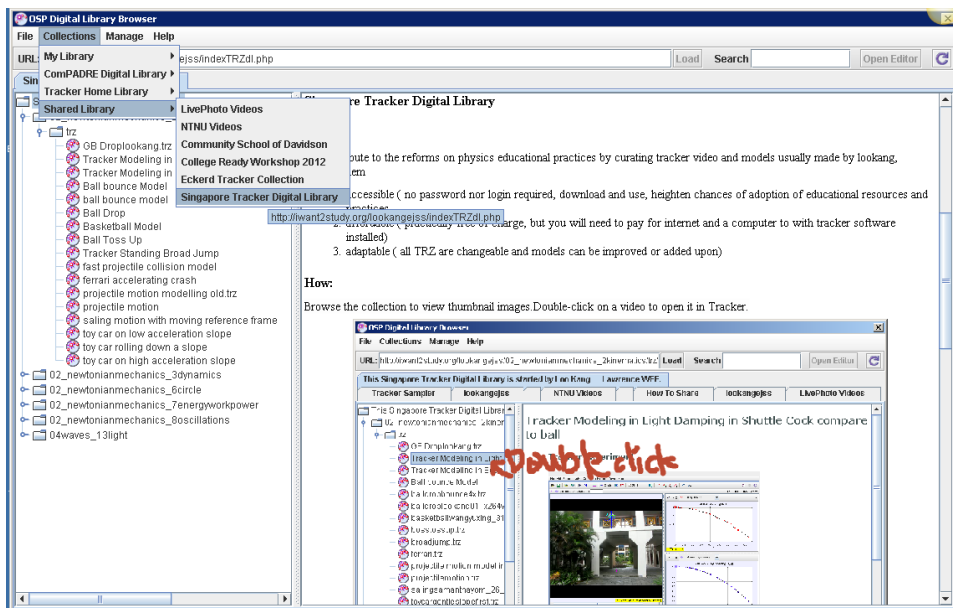

Figure 14. Officially in Tracker as Shared Library **http://iwant2study.org/lookangejss/indexTRZdl.php**

# 6   Acknowledgement

We wish to acknowledge the passionate contributions of Douglas Brown, Wolfgang Christian, Mario Belloni, Anne Cox, Francisco Esquembre, Harvey Gould, Bill Junkin, Aaron Titus and Jan Tobochnik for their creation of Tracker video analysis and modeling tool.

This research is made possible; thanks to the eduLab project NRF2013-EDU001-EL017 Becoming Scientists through Video Analysis, awarded by the National Research Foundation (NRF), Singapore in collaboration with National Institute of Education (NIE), Singapore and the Ministry of Education (MOE), Singapore.

Any opinions, findings, conclusions or recommendations expressed in this paper, are those of the authors and do not necessarily reflect the views of the MOE, NIE or NRF.

Loo Kang Lawrence WEE
Ministry of Education, Educational Technology Division, Singapore.
1 North Buona Vista Drive, Singapore 138675
Singapore
e-mail: lawrence_wee@moe.gov.sg